\documentclass[prb,twocolumn,showpacs,floatfix,amsbsy]{revtex4}
\usepackage{epsfig,color}

\begin{document}
    \author{M.P. Nowak } \affiliation{Faculty of  Physics and Applied
    Computer Science, AGH University of Science and Technology, al.
    Mickiewicza 30, 30-059 Krak\'ow, Poland}
    \author{B. Szafran} \affiliation{Faculty of Physics and Applied
    Computer Science, AGH University of Science and Technology, al.
    Mickiewicza 30, 30-059 Krak\'ow, Poland}

\title{Time dependent configuration interaction simulations
of spin swap in spin orbit coupled double quantum dots}

\begin{abstract}
 We perform  time-dependent simulations of spin exchange for an electron pair in laterally coupled quantum dots. The calculation is based on configuration interaction scheme accounting for spin-orbit (SO) coupling and electron-electron interaction
in a numerically exact way. Noninteracting electrons exchange orientations of their spins in a manner
that can be understood by interdot tunneling associated with spin precession in an effective
SO magnetic field that results in anisotropy of the spin swap.
The Coulomb interaction blocks the electron transfer between the dots but the spin transfer
and spin precession due to SO coupling is still observed.
The electron-electron interaction additionally induces
an appearance of spin components in the direction of the effective
SO magnetic field which are opposite in both dots.
Simulations indicate that the isotropy of the spin swap is restored
for equal Dresselhaus and Rashba constants and properly oriented dots.
\end{abstract}
\pacs{71.70.Gm,75.70.Tj,03.67.Lx,73.21.La}
\maketitle

\section{Introduction}
One of ideas for solid state quantum computation
employs spins of electrons confined in quantum dots \cite{lossdiv,burkard0} for information storage and processing.
This idea drives an extensive experimental research on spin manipulation,\cite{petta} detection,\cite{elzerman}, decoherence and relaxation.\cite{taylor}
Construction of a universal quantum gate requires implementation of
a controllable spin exchange operations between electrons confined in adjacent dots.
In the absence of spin-orbit (SO) coupling, the spin-spin Hamiltonian possesses an isotropic form,\cite{burkard0} i.e. depends only on relative
orientation of the spins. Anisotropic corrections \cite{kavokin,badescu,suhas} are introduced by SO coupling.
The spin processing procedures were revised \cite{bonesteel,burkard} in order to remove or minimize the anisotropy due to the SO coupling. On the other hand a practical use was invented for the anisotropic exchange interaction in construction of
 quantum gates.\cite{us,us2,us3}
A recent study \cite{baruffa} reported that the anisotropic part of the exchange interaction vanishes in zero magnetic field which should lift the limitations for spin information processing that were the major concern of the previous work.\cite{kavokin,bonesteel,burkard,us,us2,us3} The conclusion was supported by comparison \cite{baruffa} of
the double dot energy spectrum as found by the exact diagonalization technique and by a model Hamiltonian.

The spin interactions for two-electron systems are probed by charging experiments \cite{expy} that resolve the singlet-triplet
avoided crossings due to the SO coupling. These avoided crossings occur in external magnetic field, for which
anisotropy of the exchange interaction is evident. At the moment the spin dynamics for double quantum dots
in zero magnetic field can only be verified in a numerical experiment which we provide in this work.
We present results of simulation in which the spin dynamics is monitored in time.
We use the method of configuration interaction to simulate
the spin swap in laterally coupled quantum dots. Our numerically exact results indicate
that the swap process and result depend on the initial orientation of the spins
 also in zero magnetic field.
We point out that the main source of the anisotropy is the effective magnetic field due to the SO coupling \cite{meier,szafranbednarek} that leads to precession of spins of moving electrons.
The Coulomb interaction blocks the single-electron motion within the double dot. Nevertheless, the collective motion is still observed and
we find that it results in the transfer of the spin associated with its precession. Moreover, we demonstrate that the Coulomb interaction leads to an appearance
of spin components in the direction of the effective magnetic field which are opposite in both dots.

The effects of the SO coupling for electron energy spectra is  lifted
and SU(2) symmetry is restored when the
Dresselhaus and Rashba interactions acquire the same strength.\cite{schliemann} The Rashba interaction
can be controlled by external electric fields and restoration of SU(2) symmetry allows for appearance of helical spin density waves in the two-dimensional electron gas.\cite{psh}  Our simulations indicate that for equal Dresselhaus and Rashba coupling constants the spin swap becomes isotropic for carefully chosen spatial orientation of the double dot
for which the SO effective magnetic field vanishes.

\section{Theory}  We consider a two-dimensional Hamiltonian:
\begin{equation}
h=\left( \frac{\hbar^2\mathbf{k}^2}{2m^*} + V({\bf r})\right)\mathbf{1} +  H_{SIA} + H_{BIA},
\end{equation}
where $\mathbf{k}=-i\nabla$, $\mathbf{1}$ is the identity matrix, $V({\bf r})$ stands for the confinement potential, $H_{SIA}$ and $H_{BIA}$ introduce the Rashba  and Dresselhaus SO interactions, respectively.
For $x$ and $y$ axes oriented parallel to [100] and [010] crystal directions, SO interaction terms are given by $H_{BIA}= \beta \left(\sigma_x k_x - \sigma_y k_y\right) + H_{cub}$, and  $H_{SIA}=\alpha(\sigma_x k_y - \sigma_y k_x) + H_{diag}$.
The cubic Dresselhaus term $H_{cub}$ and the diagonal Rashba terms $H_{diag}$ (for their form see  [\onlinecite{my}]) are included
in the calculation but have a negligible influence on the spin evolution.
We use In$_{0.5}$Ga$_{0.5}$As effective mass $m^*=0.0465m_0$, dielectric constant $\epsilon=13.55$
and assume a model confinement potential\cite{my}
$V(x,y) = -{V_0}/{[\left( 1+ \left[ {x^2}/{R_x^2} \right]^\mu \right) \left( 1+ \left[ {y^2}/{R_y^2} \right]^\mu \right)]} + {V_{b}}/[{\left( 1+ \left[ {x^2}/{R_b^2} \right]^\mu \right) \left( 1+ \left[ {y^2}/{R_y^2} \right]^\mu \right)]}$, where  $\mu=10$,
 $V_0=50$ meV is depth of dots, $V_{b}=10$ meV is the height of the interdot barrier. The size of the double dot system in $x$ and $y$ direction
is $2  R_x=90$ nm and $2 R_y=40$ nm, respectively.  $2  R_b=10$ nm is the interdot barrier width.

Calculations for the two-electron system start by diagonalization of the Hamiltonian \begin{equation}H=h_1+h_2+\frac{e^2}{4\pi\epsilon\epsilon_0|{\bf r}_1-{\bf r}_2|}\label{ham}\end{equation} in the basis
of eigenstates of Eq. (1) that are in turn determined using a multicenter Gaussian variational wave function.\cite{my}
The eigenvalues $E_m$ of Hamiltonian (\ref{ham}) and the corresponding eigenfunctions $\Psi_m({\bf r}_1,{\bf \sigma}_1,{\bf r}_2,{\bf \sigma}_2)$ obtained from the configuration interaction scheme are used for simulation of the time evolution as described by the Schr\"odinger equation $i\hbar \frac{d\Psi}{dt}=H\Psi$,
\begin{equation}
\Psi=\sum_m c_m \exp(-iE_mt/\hbar) \Psi_m, \label{me}
\end{equation}
where $c_m$ coefficients are determined by the initial condition.
For diagonalization of Hamiltonian (\ref{ham}) we use a basis of 325 two-electron wave functions obtained from 26 lowest-energy single-electron eigenstates. The convergence of the results is discussed in the Appendix.

In order to simulate the spin swap, in the initial condition we localize the electrons in separate dots with opposite spin orientations.
We denote the initial spatial single-electron wave functions  localized in the left and right dots by
$\psi_l$ and $\psi_r$, respectively. Functions $\psi_l$ and $\psi_r$ are obtained by superposition of  bonding and antibonding orbitals
that are obtained as two lowest-energy eigenstates
of Hamiltonian (1) {\it without} the SO coupling.
The initial two-electron wave function is taken as an antisymmetrized product
$\Psi_{l\downarrow r\uparrow} = \frac{1}{\sqrt{2}}(\psi_{l}(\mathbf{r}_1)s_{+}({\sigma}_1)\psi_{r}(\mathbf{r}_2)s_{-}({\sigma}_2)-
\psi_{l}(\mathbf{r}_2)s_{+}({\sigma}_2)\psi_{r}(\mathbf{r}_1)s_{-}({\sigma}_1))$,where $s_{+}$ and $s_{-}$ are orthogonal
eigenfunctions of a chosen spin component.  Projection of these wave functions on eigenfunctions of Hamiltonian (2) defines $c_m=\langle \Psi_{l\downarrow r\uparrow}| \Psi_m \rangle$ used in Eq. (3).

Below we consider precession of a single electron spin in the effective magnetic field due to the SO coupling.\cite{meier}
For this purpose we solve the Bloch equation  \begin{equation} \frac{d\langle {\bf s}\rangle }{dt}=g\mu_b \langle {\bf B}_{SO}\times {\bf s} \rangle/\hbar,\end{equation} where $\mu_b$ is the Bohr magneton and ${\bf B}_{SO}$ is the effective magnetic field  due to the linear (dominating) terms of the SO coupling
\begin{equation} {\bf B}_{SO}= \frac{2\alpha}{g \mu_B} \left( \begin{array}{c} k_y\\ -k_x \\ 0 \end{array} \right)+\frac{2\beta}{g \mu_B} \left( \begin{array}{ccc}k_x \\-k_y \\  0\end{array} \right). \end{equation} For each time step we evaluate the righthand side of Eq. (4) using the average values as provided by the time evolution in the basis of single-electron eigenstates
\begin{eqnarray}
\left(\begin{array}{c} \frac{d\langle s_x\rangle}{dt} \\ \frac{d\langle s_y\rangle}{dt} \\ \frac{d\langle s_z\rangle}{dt}\end{array}\right) &=& \frac{2\alpha}{\hbar} \left( \begin{array}{c}
-\langle k_xs_z \rangle \\
-\langle k_ys_z \rangle \\
\langle k_x s_x \rangle + \langle k_y s_y \rangle \end{array} \right) \nonumber \\ &+&
\frac{2\beta}{\hbar} \left( \begin{array}{c}
-\langle k_ys_z \rangle \\
-\langle k_xs_z \rangle \\
\langle k_x s_y \rangle + \langle k_y s_x \rangle \end{array} \right).
\end{eqnarray}

\begin{figure}[ht!]
\epsfysize=48mm
                \epsfbox[6 275 578 580] {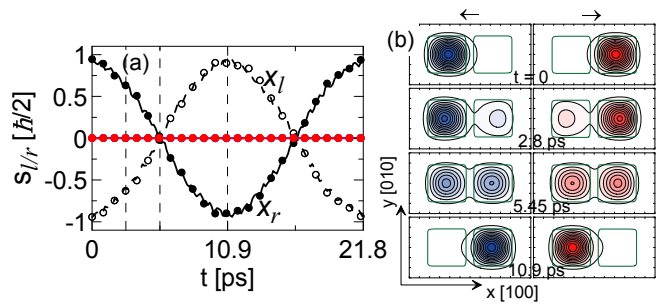}
                 \caption{
                 Spin swap simulation for quantum dots placed on the $x$ axis. In the initial condition one electron is localized in the left dot $(x<0)$ with spin oriented antiparallel to the $x$ axis and the other electron is localized in the right dot $(x>0)$ with the spin parallel to the $x$ axis. Results were
                 obtained for pure Dresselhaus coupling.
                 (a) Black lines show the $x$ component of the spin stored in the left (dashed) and right (solid) quantum dots.\cite{jak}
                 The circles indicate the results obtained without SO coupling.
                 Average values of $y$ and $z$ components is zero with (red lines) or without (red circles) SO coupling.
                                  (b) Spin-left (left column) and spin-right (right column) densities in selected moments in time. }
 \label{2e2}
\end{figure}

\begin{figure}[ht!]
\epsfysize=45mm
                \epsfbox[23 272 582 580] {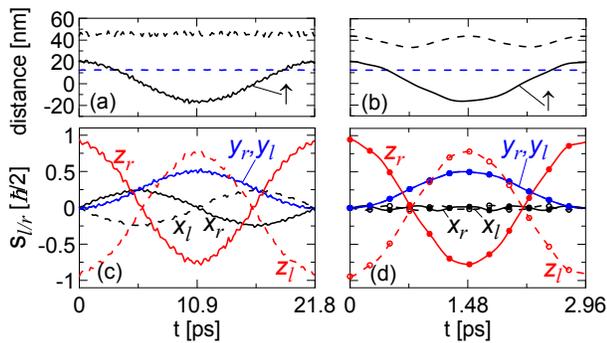}
                 \caption{
                 Simulation similar to the one presented in Fig. \ref{2e2} only for electron spins set initially parallel or antiparallel to the $z$ axis.
                 In the left (right) column of plots the Coulomb interaction is included (neglected).
                  (a,b) The average position of the spin-up density (black solid line). The electron-electron distance (dashed lines) in the $x$ (black), and $y$ (blue) directions, calculated as $\sqrt{\langle (x_1-x_2)^2\rangle}$ and $\sqrt{\langle (y_1-y_2)^2\rangle}$, respectively. (c,d) The spin components stored in the left (dashed curves) and right (solid curves) quantum dots.\cite{jak} Here and in all the other plots the $x,y$ and $z$ components of the spin are given by black, blue, and red lines, respectively. Circles in (d) -- see text.   }
 \label{2e1}
\end{figure}

\section{Results}
Two-dimensional SO coupling constants depend on the thickness $d$
of the confinement layer  $\beta=\gamma\left(\pi/d\right)^2$,
and on the value of vertical electric field $F_z$ (external or built-in)
$\alpha=\alpha_{3D}F_z$, where $\gamma$ and $\alpha_{3D}$ are bulk Dresselhaus and Rashba constants, respectively.
We use $\gamma=32.2$ meVnm$^{3}$ for the InGaAs alloy \cite{saikin} and assume $d=5.4$ nm
which gives  $\beta=10.8$ meV nm. This seems a maximal value of the coupling constant that can be practically achieved in a InGaAs quantum dot.
The bulk Rashba constant of the alloy is $\alpha_{3D}=0.572$ nm$^2$ (after Ref. [\onlinecite{silva}]). The electric field needed to produce  $\alpha=10.8$ meV nm is 18.9 meV/nm, which is equivalent to the confinement potential drop of 102 meV across the dot of the height $d=5.4$ nm.
Results presented below stay qualitatively unaffected for smaller values of the coupling constants or weaker interdot coupling (for the latter -- see the Appendix).

\subsection{Quantum dots placed along $[100]$ direction}
Let us first assume that the confinement potential is symmetric in the growth direction ($\alpha=0$) and that the centers of the dots are placed on the $x$ axis.
Initially the spin in the left (right) dot is set parallel (antiparallel) to the $x$ axis.
Figure \ref{2e2}(a)  shows the time dependence of the average spin stored in the left and right dot.\cite{jak}
For $t=0$ the spins in the left and right dots are not exactly equal to $\pm {\hbar/2}$ due to leakage of $\psi_l$ ($\psi_r$) functions to the right (left) dot [see Fig. \ref{2e2}(b) for $t=0$].
The spin swap is completed at $t=t_s=10.9$ ps and no other component of the spin is generated during the process.

The swap process {\it exactly} as illustrated in Fig. \ref{2e2}(a) is obtained
in the absence of SO interaction independent of the choice of the direction
in which the electron spins are initially set antiparallel to each other.
In order to present the effects of SO coupling for the spin swap, let us change the initial condition.
At $t=0$ the spin in the right (left) dot is now set parallel (antiparallel) to the $z$ axis [see Fig. \ref{2e1}(c)].
At $t=t_s$ the absolute values of $\langle s_z\rangle$ are visibly reduced as compared to the initial condition and the
spins in both dots acquire an identical non-zero value of the $y$ component. Moreover, opposite components of the spin in the $x$ direction
are generated in both dots during the swap. The $x$ spin components are maximal at $t=t_s/2$ and disappear once the swap is completed.

\begin{figure}[ht!]
\epsfysize=43mm
                \epsfbox[24 290 578 566] {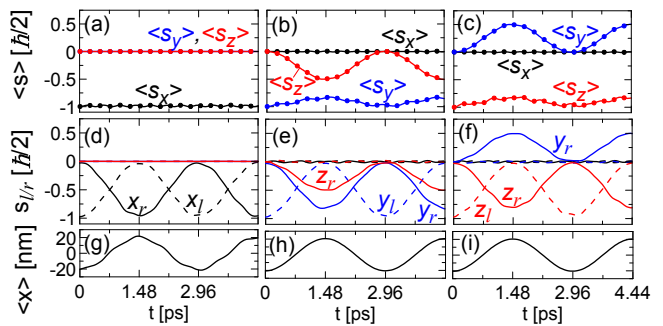}
                 \caption{Simulation for a single electron, pure Dresselhaus coupling and dots placed on the $x$ axis. The electron is initially localized in the left dot with spin oriented antiparallel to the $x$, $y$ and $z$ axis in the left, central and right columns of plots, respectively. (a,b,c) Average values of spin components are given by curves. The circles indicate the results obtained from the Bloch equation (4). (d,e,f) The spin stored in the left (dashed lines) and the right (solid lines) quantum dots. Black, blue and red lines show the results for $x$, $y$ and $z$ components, respectively.
                 (g,h,i) The average $x$ position of the electron packet.  }
 \label{1e2}
\end{figure}

For $\alpha=0$
the SO effective magnetic field of Eq. (5) is oriented along the $x$ axis -- the direction of electron tunneling between the dots.
In order to evaluate the effects of the spin precession we performed
calculations for a single electron. We localize the electron in the left dot and assume that its spin is oriented antiparallel to the $x$ [Fig. \ref{1e2}(a,d,g)],
$y$ [Fig. \ref{1e2}(b,e,h)], and $z$ axes [Fig. \ref{1e2}(c,f,i)]. The electron wave packet oscillates between the left and right dots [Fig. \ref{1e2}(g,h,i)]. The electron -- localized initially in the left dot with the spin aligned with the $z$ ($y$) direction -- acquires
a non-zero $y$ ($z$) component of the spin when it tunnels to the right dot [Fig. \ref{1e2}(e) and \ref{1e2}(f)].
Figures \ref{1e2}(a,b,c) show that the spin evolution as obtained by integration of the Bloch equation (4) describing the spin precession
perfectly agrees with the results of the main simulation.
For the spin initially set antiparallel to the $x$ axis -- i.e. aligned with ${\bf B}_{SO}$ --
no precession is observed [Fig. \ref{1e2}(a,d)].

Let us now return to the problem of two electrons with spins initially  parallel and antiparallel   to the $z$-axis.
The results for noninteracting pair of electrons [given by lines in Fig. \ref{2e1}(d)] are exactly reproduced
by the sum of single-electron solutions [dots in Fig. \ref{2e1}(d)].
Figure \ref{2e1}(b) shows that the horizontal electron-electron distance oscillates as the noninteracting electrons tunnel from one dot to the other
and meet at the interdot barrier. The same $y$ component of the spin is generated in both dots in any moment of time. The electron spins are initially oppositely oriented with respect to the $z$ axis and move in the opposite directions along the $x$ axis. Accordingly, the change of $y$ spin component as given by Eq. (6) has the same sign for both electrons.

When the Coulomb repulsion is switched on, the oscillation of the electron-electron distance is no longer observed [Fig. \ref{2e1}(a)] -- the electrons occupy fixed positions in separate dots. The electron density -- the sum of spin-up and spin-down electron densities -- is nearly stationary,
but the components of the sum are not.
In Fig. \ref{2e1}(a,b) with the black solid line we plotted the center-of-mass of the spin-up electron density packet.
We notice that this center oscillates in a very similar way for both interacting and noninteracting electrons. Also the spin swap as obtained for interacting electrons is similar to the one found in the absence of the Coulomb repulsion [cf. Fig. \ref{2e1}(c) and (d)], only the swap time  is increased by a factor of 10 as the Coulomb repulsion enhances the effective height of the interdot barrier. The only qualitative feature introduced by the Coulomb interaction is the noticeable  oscillation of the $x$ component of the spin.
We found as a general rule for interacting electrons that during the spin precession opposite spin components in the direction of ${\bf B}_{SO}$ appear in both dots.

\begin{figure}[ht!]
\epsfysize=34mm
                \epsfbox[23 312 579 525] {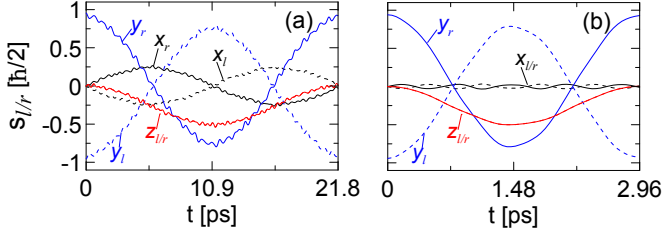}
                 \caption{(a) Same as Fig. \ref{2e1}(c). (b) Same as Fig. \ref{2e1}(d) only for
                 electron spins initially antiparallel in the $y$ direction.
                  }
 \label{2eD}
\end{figure}

 Figure \ref{2eD} shows the results for spins initially antiparallel in the $y$ direction, still for the pure Dresselhaus coupling. The appearance of the $z$ component of the spin during the swap is due to spin precession and is observed for both interacting [Fig. \ref{2eD}(a)] and noninteracting [Fig. \ref{2eD}(b)] electrons. For the Coulomb interaction present the
opposite spin components in the $x$ direction (aligned with ${\bf B}_{SO}$) appear in the dots, similarly as presented in Fig. \ref{2e1}(c) for the spins initially aligned with the $z$ axis.

\begin{figure}[ht!]
\epsfysize=70mm
                \epsfbox[26 171 567 672] {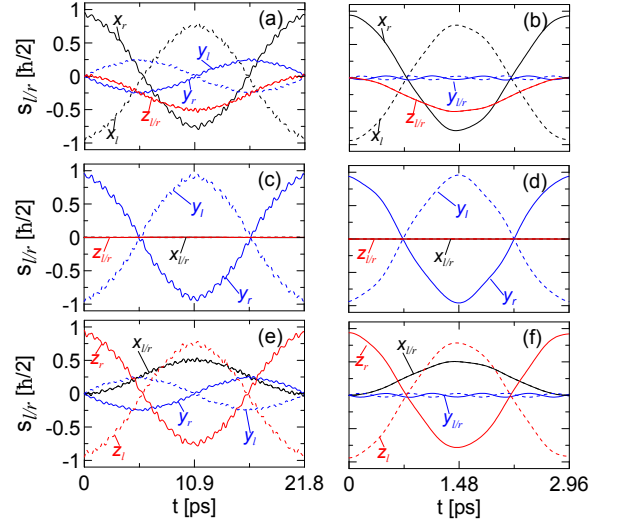}
                 \caption{Simulation for pure Rashba coupling and centers of dots placed on the $x$ axis. The electrons in the left and right dots initially possess opposite spins in $x$ (a,b), $y$ (c,d), and $z$ (e,f) directions. Black, blue and red curves show the $x$, $y$ and $z$ spin components stored in the left (dashed curve) and the right (solid curve) dots. Plots (b,d,f) were obtained for neglected electron-electron interaction which is included in (a,c,e).}
 \label{2eR}
\end{figure}

\begin{figure}[ht!]
\epsfysize=70mm
                \epsfbox[26 171 567 672] {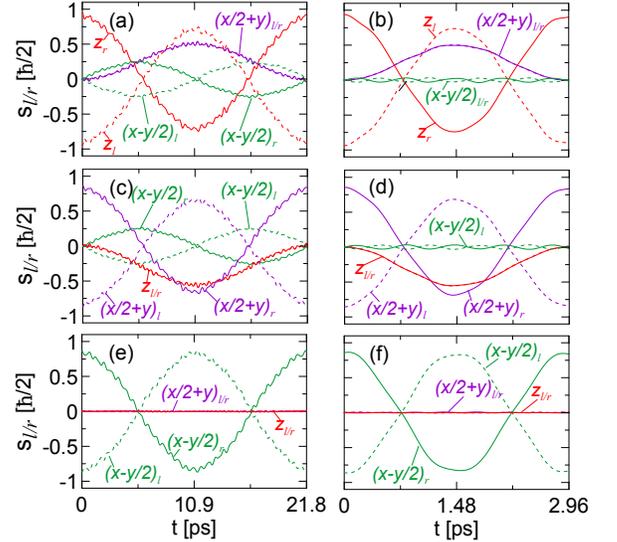}
                 \caption{Simulation for both SO coupling types present $\alpha=\beta/2=5.4$ meV nm and centers of dots placed on the $x$ direction. The electrons in the left and right dots initially possess opposite spins along $z$ ([001]) direction  in (a,b), $x/2+y$ ($[1,2,0]$) in (c,d), and $x-y/2$ ($[2,\overline{1},0]$) in (e,f). Purple, green and red curves show the $[1,2,0]$, $[2,\overline{1},0]$ and $[0,0,1]$ spin components stored in the left (dashed curve) and the right (solid curve) dots. Plots (b,d,f) were obtained for neglected electron-electron interaction which is included in (a,c,e).}
 \label{apb}
\end{figure}

To further illustrate the findings of the above paragraph let us consider the case of pure Rashba  coupling
(Fig. \ref{2eR}).
In III-V material the Dresselhaus coupling can not be completely removed. Nevertheless, it can be small as compared to
the Rashba coupling provided that the dots height is large and / or strong electric field is applied in the growth direction.
For the pure Rashba coupling and the considered orientation of the dots  ${\bf B}_{SO}$ is aligned with the $y$ axis [see Eq.(5)].
For the spins initially parallel and antiparallel to this axis, the spin swap  [Fig. \ref{2eR}(c,d)] occurs without generation
of neither $x$ nor $z$ spin components. For the spins initially aligned  with $x$ ($z$) axis one
observes appearance of $z$ ($x$) spin component -- the same in both dots --  for both
interacting and noninteracting electrons -- see Fig. \ref{2eR}(a,b) [Fig. \ref{2eR}(e,f)] that results from the spin precession.
According to Eq. (6) for $\beta=0$ the electron spin that is initially parallel to the $x$ axis and moves in the $x$ direction acquires the $z$ component of the same sign as the spin antiparallel to the $x$ axis that moves in the $-x$ direction [Fig. \ref{2eR}(a,b)].
When the electron-electron interaction
is present we additionally observe [see Figs. \ref{2eR}(a) and \ref{2eR}(e)] an appearance of opposite spin components
in the $y$ direction (i.e. direction of ${\bf B}_{SO}$)
in both dots.

Finally, Fig. \ref{apb} shows the results for both coupling types present with $\beta=10.8$ meV nm and $\alpha=5.4$ meV nm.
For the electrons moving between the two dots, the SO magnetic field vector is aligned with the $x-y/2$ line,
i.e. the $[2\overline{1}0]$ crystal direction. For the discussion of the spin behavior we consider this crystal direction
and two orthogonal ones: within the plane of confinement [120], and in  the growth direction [001].
 When the electrons spins are initially aligned with the SO field vector, the swap
goes without generation of other components of the spin [Fig. \ref{apb}(e,f)].
For spins initially antiparallel to the  $[001]$ direction [Fig. \ref{apb}(a,b)] the precession of the spins
leads to an appearance of  $[120]$ spin component and vice versa [Fig. \ref{apb}(c,d)].
The electron-electron interaction for both initial spin orientations that are orthogonal to ${\bf B}_{SO}$
leads to appearance of the spin components parallel to the ${\bf B}_{SO}$ direction [see the green curves in Fig. \ref{apb}(a) and (e)]
in the form noticed above for the pure Dresselhaus and Rashba coupling: vanishing at $t=t_s$, maximal at $t=t_s/2$, and
always opposite in both dots.

\subsection{Spin oscillations due to the electron-electron interaction}
The results presented above indicate that during the  swap of spins that are initially perpendicular to ${\bf B}_{SO}$, the electron-electron interaction induces appearance of  spin components in the direction of this vector that are opposite in both dots  [Fig. 2(c), Fig. 4(a), Fig. 5(a,e), Fig. 6(a,b)].
The net spin in the direction of  ${\bf B}_{SO}$ remains  zero, in contrast to the spin generated in the direction perpendicular to ${\bf B}_{SO}$ by the spin precession. The spin components in  ${\bf B}_{SO}$ direction vanish at the end of the swap $t=t_s$, however
they are maximal at $t=t_s/2$.
Note that the XOR gate employs the square-root-of-a-swap operation,\cite{burkard0}
i.e. the spin swap process interrupted exactly at $t_s/2$.

Let us analyze the background of the appearance of spin components in the direction of ${\bf B}_{SO}$ vector.
We focus on the simulation presented in Fig. 2(c), for which  $\alpha=0$, the dots are placed along the $x$ axis hence ${\bf B}_{SO}$ is aligned with the $x$ axis,
and the spins are initially set parallel and antiparallel to the $z$ axis.

\begin{figure}[ht!]
\epsfysize=70mm
                \epsfbox[20 180 570 663] {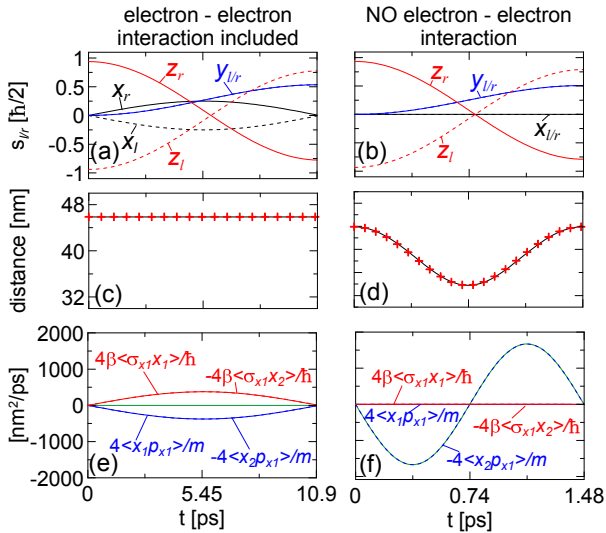}
                 \caption{(a,b) same as Fig. \ref{2e1}(c,d) but for the basis limited to four lowest-energy two-electron states. (c,d) Electron-electron distance ($\langle(x_1-x_2)^2\rangle^{1/2}$) as calculated for the wave function given by Eq. (4)  (black line) and integrated using Eq. (7) (red crosses)
                 starting from the initial condition. (e,f) Real parts of righthand side terms of Eq. (7).}
 \label{ehre}
\end{figure}

By the Ehrenfest theorem the average electron-electron distance in the $x$ direction changes in time as \begin{eqnarray}
\frac{d}{dt} \langle (x_1 - x_2)^2 \rangle &=& \frac{4}{m^*}\langle x_1p_{x1}-x_2p_{x1}\rangle \nonumber \\ &+& \frac{4\beta}{\hbar}\langle\sigma_{x1}x_1 - \sigma_{x1}x_2 \rangle - \frac{2i\hbar}{m^*},
\end{eqnarray}
where the last term compensates for the imaginary part due to the non-Hermitian $x_1p_{x1}$ operator.
The electron-electron distance as presented in Fig. 2(a) is nearly constant but contains a rapid oscillation of small amplitude which
results from a difference between the electron-electron separation in the initial condition and the equilibrium distance for interacting electrons (see Appendix).
For the purpose of analysis of Eq. (7) it is useful to limit the basis used for Eq. (3) to 4 lowest-energy two-electron states,
which correctly describes the spin evolution (see Appendix) but is free of this rapid oscillation.
The results for the spin evolution, electron-electron distance and  the right-hand-side terms of Eq. (7) are displayed in Fig. \ref{ehre}.
For noninteracting electrons, at the right-hand side of Eq. (7) only the $\langle x_2p_{x1}\rangle=\langle x_2 \rangle \langle p_{x1}\rangle$ term has a non-zero real part [Fig. \ref{ehre}(f)], which oscillates due to independent tunneling of electrons which go from one dot to the other with periodically changing positions and momenta. On the other hand the average value of  $x_1 p_{x1}$ operator is purely imaginary.\cite{pi}
      For noninteracting electrons the term of Eq. (7) containing $x\sigma_x$ vanishes [Fig. 7(f)] and so does $\sigma_x$ in both dots [Fig. 7(b)]. Thus the oscillation of the electron-electron distance observed in Fig. \ref{ehre}(d) is only due to the mean value of $x_2p_{x1}$ operator.

\begin{figure}[ht!]
\epsfysize=70mm
                \epsfbox[12 118 589 669] {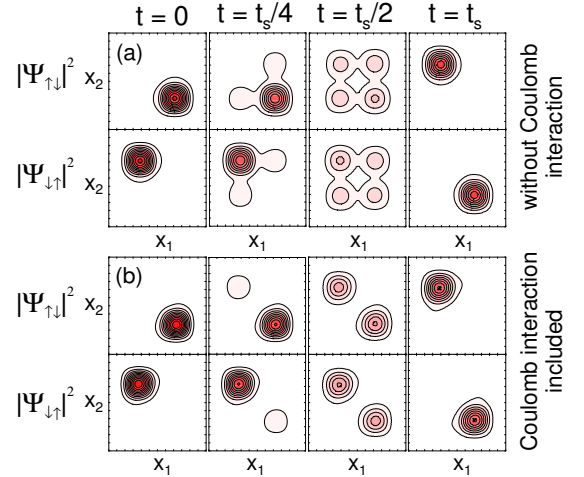}
                 \caption{Probability densities for the components of the two-electron wave functions corresponding to opposite
                 spin orientations plotted on $x_1,x_2$ plane along the axis of the system $y_1=y_2=0$ for chosen moments in time during the spin swap. Parameters are the same as in Fig. \ref{ehre}. The spins are initially oriented parallel and antiparallel to the $z$ axis for noninteracting (a) and interacting electrons (b). The spin swap time is $t_s=1.48$ ps for noninteracting electrons (a) and $t_s=10.9$ ps for the interacting ones (b).}
 \label{x1x2}
\end{figure}

Interacting electrons keep their relative distance [Fig. \ref{ehre}(c)], so the terms at the righthand side of Eq. (7) must cancel one another.
They indeed do [Fig. \ref{ehre}(e)]. Remarkably, in contrast to the case without Coulomb interaction [Fig. \ref{ehre}(f)], for interacting electrons one finds [Fig. \ref{ehre}(e)] \begin{equation}\langle x_1 p_{x1}\rangle=-\langle x_2 p_{x1}\rangle\end{equation}
and \begin{equation}\langle \sigma_{x1} x_1\rangle=-\langle \sigma_{x1} x_2\rangle.\end{equation}
 Relations (8) and (9) can be explained by analysis of the electron motion which becomes collective when the electron-electron interaction is introduced. The two-electron wave function of Eq. (3),
can be written as a four-component wave function \cite{szafrannowak} $\Psi_1=\Psi_{\uparrow\uparrow}({\bf r}_1,{\bf r}_2)$, $\Psi_2=\Psi_{\uparrow\downarrow}({\bf r}_1,{\bf r}_2)$,  $\Psi_3=\Psi_{\downarrow\uparrow}({\bf r}_1,{\bf r}_2)$, and $\Psi_4=\Psi_{\downarrow\downarrow}({\bf r}_1,{\bf r}_2)$, each corresponding to a given direction of the spin for a given electron label
[e.g. $\Psi_{\uparrow\downarrow}({\bf r}_1,{\bf r}_2)$ corresponds to electron of position ${\bf r}_1$ (${\bf r}_2$) with spin oriented parallel (antiparallel) to the $z$ axis].
Antisymmetry of the wave function with respect to the electron interchange implies $\Psi_1({\bf r}_1,{\bf r}_2)=-\Psi_1({\bf r}_2,{\bf r}_1)$,
$\Psi_4({\bf r}_1,{\bf r}_2)=-\Psi_4({\bf r}_2,{\bf r}_1)$, and $\Psi_2({\bf r}_1,{\bf r}_2)=-\Psi_3({\bf r}_2,{\bf r}_1)$.
For the exchange of initially opposite spins $\Psi_3$ and $\Psi_4$ components are most relevant.
Snapshots of $|\Psi_3|^2$ and $|\Psi_4|^2$ are displayed in Fig. \ref{x1x2} as functions of $x_1$ and $x_2$ calculated along the axis of the double dot $y_1=y_2=0$ for noninteracting [Fig. \ref{x1x2}(a)] and interacting [Fig. \ref{x1x2}(b)] electrons.
Figure \ref{x1x2} shows that in the initial condition  the electrons occupy separate dots and that the spin contained in the right (left) dot is oriented parallel (antiparallel) to the $z$ direction.
Spin orientation is inverted after the swap ($t=t_s$).
For noninteracting electrons at $t=t_s/4$ and $t=t_s/2$ we notice [Fig. \ref{x1x2}(a)]
that probabilities to find both electrons in the same dot (i.e. on the diagonal $x_1=x_2$ of the plots) is non-zero,
which results from an independent electron tunneling between the dots.
Without the Coulomb interaction
the spin swap occurs as due to single-electron tunneling.

The interacting electrons do not occupy the same dot [see the vanishing probability density on the diagonal of plots presented in Fig. 8(b)] and the single-electron interdot tunneling is blocked. The interdot tunneling of separate
wave function components is still observed [Fig. 8(b) for $t=t_s/4$ an $t=t_s/2$] but it occurs only along the antidiagonal of the plot $x_2=-x_1$.
Therefore, one can replace $x_2$ by $-x_1$ in the righthand side of Eqs. (8) and (9) which explains these relations.
For the electron-electron distance to be unchanged the terms $\langle x_1p_{x1} \rangle $ and $\langle -x_2p_{x1}\rangle$ of Eq. (7)
need to be canceled by the terms that contain the $x$ component of the spin.
The operator $x_1 \sigma_{x_1}$ produces a non-zero contribution since the $x$ spin component generated in the left dot ($x<0$)
has opposite sign [see Fig. 7(a)] than the one generated in the right dot ($x>0$).
We conclude that the generation of opposite spin components in the direction of the effective magnetic field
is a consequence of fixed electron-electron distance and collective evolution of the two-electron wave function that are both
induced by the Coulomb interaction.

\begin{figure*}[ht!]
\epsfysize=50mm
                \epsfbox[20 340 578 490] {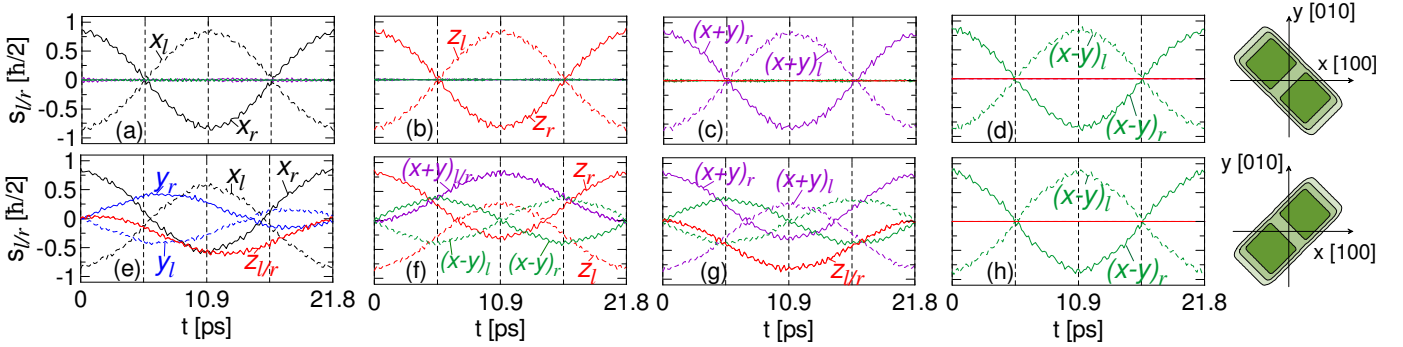}
                 \caption{Spin stored in separate dots for two interacting electrons with equal linear Dresselhaus and Rashba constants $\alpha=\beta=10.8$ meV nm.      The first and second row of plots correspond to different alignments
                  of the double dot as shown schematically at the right end of the figure.
                 In (a,e) (b,d) the spins are initially set parallel or antiparallel to the
                  $x$, and $z$ axes, respectively.  In (c,g) the spins are initially set in the $x+y$ (i.e. [110]) direction -- within the plane of confinement and perpendicular to ${\bf B}_{SO}$ direction ($[1\overline{1}0]$). In (d,h) the spins are initially aligned
                  with ${\bf B}_{SO}$ vector.
              The dashed (solid) lines show the results for the left (right) dot. Black, blue, red, purple and green colors correspond to $x$, $y$,  $z$, $x+y$ and $x-y$ components of the spin, respectively.
                  }
                  \label{2e4}
\end{figure*}

\subsection{Spin exchange for $\alpha=\beta$}

For $\alpha=\beta$ the linear SO terms of Hamiltonian (1) commute \cite{schliemann} with  $\frac{1}{\sqrt{2}}(\sigma_x-\sigma_y)$ operator
and the effective magnetic field   ${\bf B}_{SO}=\frac{2\alpha}{g\mu_b}(k_x+k_y)(1,-1,0)^T$ is aligned with $[1\overline{1}0]$ crystal direction.  We performed simulations of the spin swap for  $\alpha=\beta=10.8$ meV nm. For the dots aligned in the $x$ direction that were discussed
  above the results are qualitatively identical to the ones presented in Fig. 6 only with redefined direction of ${\bf B}_{SO}$ vector.
For $\alpha=\beta$ the direction of ${\bf B}_{SO}$  does not depend on the orientation of the dots, however the strength of this field does.
For the diagonal ($[110]$) orientation of the dots (the lower row of plots in Fig. \ref{2e4}) the electrons tunnel between the dots with equal $k_x$ and $k_y$ wave vectors, so the field should be relatively the strongest. On the other hand for the quantum dots oriented along the $[1\overline{1}0]$ direction (the upper row of plots in Fig. \ref{2e4}) $k_x$ and $k_y$ have opposite sign for the electrons tunneling between the dots, so ${\bf B}_{SO}$ should be expected to vanish.

The simulations show that spin exchange occurs in the same manner for the diagonal ($[110]$) and antidiagonal ($[1\overline 10]$) orientation of the dots
only when the spins of electrons are initially antiparallel in the direction of  ${\bf B}_{SO}$ ($[1\overline 10]$ or $x-y$) -- see Fig. 9(d) and 9(h). For the diagonal orientation of the dots and initial alignment of the spins in the direction perpendicular to the ${\bf B}_{SO}$ vector:
in $z$ direction [Fig. 9(f)] and in $(x+y)$  direction [Fig. 9(g)] one observes generation of $(x+y)$ and $z$ spin components, respectively. The precession is accompanied by generation of opposite spin components in ${\bf B}_{SO}$ direction [Figs. 9(f,g)] in consistence
with the results discussed above.

Note, that the anisotropy of the swap for the diagonal orientation of the dots is observed in spite of the fact that for $\alpha=\beta$ the energy spectrum of SO-coupled system is identical \cite{schliemann} to the one obtained in the absence of SO coupling. For $\alpha=\beta$ the SO coupling
does not affect the energy spectrum at zero magnetic field but the effective SO magnetic field is still present.

For the antidiagonal ($[1\overline{1}0]$) orientation of the dots (the upper row of plots in Fig. \ref{2e4}) the spin swap becomes perfectly isotropic and occurs in the same manner independent of the initial spin orientation.
Not a trace of spin precession is present in accordance with the single-electron picture of the electron tunneling
that goes with $k_x=-k_y$ in any moment in time  which implies ${B}_{SO}=0$.

\subsection{Discussion}

 The
present study indicates that the spin swap as originally defined for the purpose of controllable coupling of spin qubits\cite{burkard0} localized in separate quantum dots is generally anisotropic when the spin-orbit coupling is present.
The anisotropy of the spin swap results
from the effective magnetic field due to the spin-orbit coupling. This field changes the direction
of the electron spin as it moves in space.
The study of Ref. \onlinecite{baruffa} indicated that
for a carefully chosen computational two-electron basis the exchange Hamiltonian becomes formally isotropic at zero magnetic field.  The proposed \cite{baruffa} basis is obtained by a unitary transformation of a separable basis of singlet and triplet states. The unitary transformation [Eq. (13) of Ref. \onlinecite{baruffa}] produces basis elements
in which the spin and spatial degrees of freedom are entangled, i.e. direction of the electron spin depends on its position in space.
For the purpose of the quantum computation any basis can in principle be chosen. However, the entangled basis that allows for a simpler form of the Hamiltonian requires a more challenging handling of the quantum information, which in fact should be stored by entangled
spin-orbital wave functions rather than by the electron spin itself.
The practical usage of the entangled basis calls for new physical procedures for preparation of the initial state and read out of the quantum computation result.

For construction of the universal quantum gate the two-spin operations need to be complemented by single spin rotations.
The direct idea to perform the latter is to put the system in external magnetic field to split the spin states of the single-electron
and exploit the Rabi oscillations in resonant radiation of microwave or radio frequency.\cite{kuw}
In presence of the external magnetic field (${\bf B}$) the spin swap becomes anisotropic even without SO coupling,
since the electron spins precess in ${\bf B}$ unless
they are initially aligned with the direction of the external field.
In order not to interfere with the spin exchange the single-spin rotations should be applied without the external magnetic field.
The original idea for that purpose was the electrically controlled coupling of a selected spin to a ferromagnetic medium.\cite{lossdiv}
It was also demonstrated that the single-spin rotations can be performed using the spin precession in the SO effective magnetic field which occurs
 when the electron is made to move, e.g. along closed trajectories.\cite{noe} This idea for the single-spin rotations\cite{noe} does not
 require the external magnetic field or irradiation.

\section{Summary and Conclusions}
We presented numerically exact simulations of the spin swap for two-electron SO-coupled double quantum dots.
The study covered  Dresselhaus and Rashba interactions and various spatial orientations of the double dot.
The swap of spins as observed in time-dependent simulation involves four mechanisms: (i) direct tunneling which consists in electron carrying its spin from one dot to the other
 (ii) the spin tunneling which still occurs when the direct tunneling is blocked by the electron-electron interaction (iii) the precession of the spin moving in the effective magnetic field due to the spin-orbit coupling and (iv) generation of opposite spin components in the direction of the effective magnetic field which is observed for interacting pair of electrons.
  The third and fourth mechanisms of the above list can be switched off for initial
spin orientation aligned with the effective magnetic field vector ${\bf B}_{SO}$. For the initial orientation of the spins in one of the directions perpendicular to ${\bf B}_{SO}$ the spin in the other perpendicular direction is generated during the swap as a consequence of the spin precession.
We argued that mechanism (iv)
is necessary to maintain a constant electron-electron distance and is a consequence of a collective motion of the electrons within the inner degrees of freedom which is still observed when the single-electron tunneling between the dots is blocked by the Coulomb repulsion.
We also demonstrated that for both coupling types present the spin swap is largely affected by a specific orientation of the double dot system within the (001) plane of confinement
via the strength of the effective magnetic field. In particular, we demonstrated that the SO coupling effects can be eliminated from the spin swap process for quantum dots aligned with $[1\overline{1}0]$ crystal direction. For this orientation of the dots
and the Rashba constant tuned to match the Dresselhaus constant ${\bf B}_{SO}$ vanishes and the spin swap becomes perfectly isotropic.

{\it Acknowledgements.}
This work was supported by MPD Programme
"Krakow Interdisciplinary PhD-Project in Nanoscience and Advanced Nanostructures''
 operated within the Foundation
for Polish Science and co-financed by the EU European Regional
Development Fund  and by a research project N N202 103938
supported by Ministry of Science an Higher Education (MNiSW) for 2010-2013.
 Calculations were performed in
    ACK\---CY\-F\-RO\-NET\---AGH on the RackServer Zeus.

\begin{figure}[ht!]
\epsfysize=50mm
                \epsfbox[12 255 590 576] {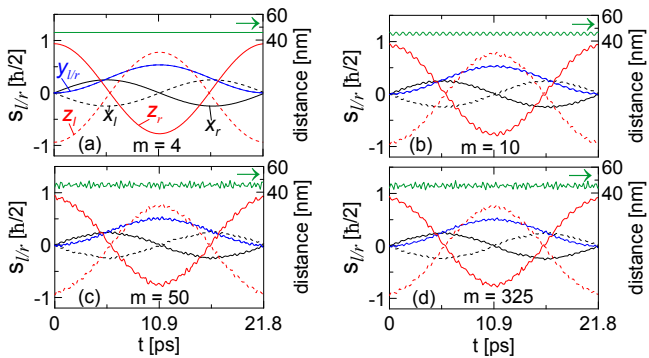}
                 \caption{Results of simulations for the pure Dresselhaus coupling and dots placed on the $x$ axis for the
                 interacting electrons with spins initially oriented antiparallel in the $z$ direction. Black, blue and red curves show the spin components stored in the left (dashed curve) and the right (solid curve) dots. The green curve at the top of the plot
                 is the average electron-electron distance that is referred to the right axis.
                 The results presented in (a), (b), (c) and (d) correspond to the two-electron basis containing $m=4, 10, 50$ and 325 lowest-energy
                 eigenstates of Hamiltonian (2).}
 \label{2ezb}
\end{figure}

\begin{figure}[ht!]
\epsfysize=30mm
                \epsfbox[12 255 590 576] {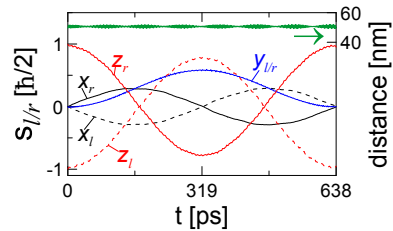}
                 \caption{Same as Fig. \ref{2ezb}(d) only for interdot barrier height increased from $V_b=10$ meV to $V_b=50$ meV. }
 \label{2evb}
\end{figure}

\section*{Appendix}
This Appendix presents convergence of the results with respect to the energy cut-off used in Eq. (3).
Let us consider the dots placed on the $x$ axis and electron spins initially oriented parallel and antiparallel to the $z$ axis
that was discussed in the context of Fig. 2(c). Fig. 10 shows the results for $m=4$, 10, 50 and 325 two-electron lowest-energy eigenstates used as basis elements in Eq. (3).
The basis with $m=4$ covers the ground-state and three-fold degenerate excited state \cite{my} --  which in the absence of SO coupling corresponds to the spin triplet. The energy separation of the ground-state
and the excited state is $\Delta E=0.189$ meV, which well corresponds to the spin swap time\cite{burkard0} $t_s=\pi/\Delta E=10.9$ ps. The basis of 10, 50 and 325 elements covers all the two-electron eigenstates of the energy that exceeds the ground-state energy by not more than  8.9, 19.4 and 65.2 meV, respectively.
For $m>4$ a rapid and low amplitude oscillation appears in the results of Fig. 10.
This oscillation results from a difference between the electron-electron separation in the initial condition and the equilibrium distance for interacting electrons. The initial condition is taken from single-electron wave function obtained for noninteracting electrons which are projected onto the basis
of two-electron eigenstates (see Section II). The electron-electron distance is nearly the same in the lowest-energy four states (singlet and triplet states). The constant electron-electron distance obtained for $m=4$ [Fig. 10(a)] is the equilibrium distance for interacting electrons in the ground-state. For larger $m$ the basis resolves between the equilibrium distances for interacting and noninteracting electrons, hence the appearance of the rapid oscillations of electron-electron distance and the resulting oscillations of the spins. The oscillations do not affect the mechanism of the spin swap nor the swap time
and have a small amplitude which can be further reduced by a choice of confinement parameters. In particular Fig. 11 shows the results for the barrier height increased from 10 do 50 meV.\cite{my} The Coulomb interaction affects weakly the electron-electron equilibrium distance and the rapid oscillations have a negligibly small amplitude also for $m=325$.
The results presented in this paper were obtained for the fully convergent 325 element basis with the exception of Subsection III.B, where we use the four-element basis for simplicity.  For $\Delta E=65.2$ meV the shortest oscillation period that can be accounted for is 0.06 ps.


\begin{thebibliography}{00}
\bibitem{jak} The spin component $i$ stored in the left dot is calculated as:
$s_{l}=\langle\Psi(\mathbf{r_1},{\bf \sigma}_1, \mathbf{r_2}, {\bf \sigma}_2)|s^i_1\kappa_{l}(\mathbf{r_1}) + s^i_2\kappa_{l}(\mathbf{r_2})|\Psi(\mathbf{r_1},{\bf \sigma}_1,\mathbf{r_2},{\bf \sigma}_2)\rangle$,
where $s^i$ is the spin component $i$ operator and
$\kappa_l(x)$ is the step function that takes value 1 in the left dot and 0 in the right dot. Formula for the spin stored
in the right dot $s_r$ is the same but with $\kappa_l$ replaced by $\kappa_r(x)=1-\kappa_l(x)$, and $s_l+s_r=\langle s^i_1+s^i_2\rangle$.
\bibitem{pi} This can be explained as follows.  The two-electron problem without the Coulomb interaction separates into two single-electron ones.
 According to the present approach initially the spatial single-electron wave function  is a superposition of the ground-state wave function of even parity $\psi_0$ and the first-excited state wave function of odd parity $\psi_1$. One can arrange for $\psi_0$ and $\psi_1$ to be real-valued.
  In the evaluation of the righthand side of Eq. (7) for noninteracting electrons the cross terms $\langle \psi_0 |xp_x |\psi_1\rangle$ vanish for the parity reason, and $\langle xp_x \rangle$ is purely imaginary for any localized real wave function.
\end{thebibliography}
\end{document}